\pdfoutput=1

\documentclass[12pt,a4paper]{article}

\usepackage{ifthen} 
\newboolean{pdflatex}
\setboolean{pdflatex}{true} 

\newboolean{articletitles}
\setboolean{articletitles}{true} 

\newboolean{uprightparticles}
\setboolean{uprightparticles}{false} 

\def\paperauthors{LHCb collaboration} 
\def\paperasciititle{First observation of excited Omegab states} 
\def\papertitle{First observation \\ of excited~$\Omegab$~states} 
\def\paperkeywords{{High Energy Physics}, {LHCb}} 
\def\papercopyright{\the\year\ CERN for the benefit of the LHCb collaboration} 
\def\paperlicence{CC-BY-4.0 licence}
\def\paperlicenceurl{https://creativecommons.org/licenses/by/4.0/}

\usepackage[top=1in, bottom=1.25in, left=1in, right=1in]{geometry}

\usepackage{microtype}
\usepackage{lineno}  
\usepackage{xspace} 
\usepackage{caption} 

\usepackage{graphicx}  
\usepackage{color}
\usepackage{colortbl}
\graphicspath{{./figs/}} 
\DeclareGraphicsExtensions{.pdf,.PDF,png,.PNG}

\usepackage{amsmath} 
\usepackage{amssymb}
\usepackage{amsfonts}
\usepackage{upgreek} 

\newcommand*\patchAmsMathEnvironmentForLineno[1]{%
\expandafter\let\csname old#1\expandafter\endcsname\csname #1\endcsname
\expandafter\let\csname oldend#1\expandafter\endcsname\csname
end#1\endcsname
 \renewenvironment{#1}%
   {\linenomath\csname old#1\endcsname}%
   {\csname oldend#1\endcsname\endlinenomath}%
}
\newcommand*\patchBothAmsMathEnvironmentsForLineno[1]{%
  \patchAmsMathEnvironmentForLineno{#1}%
  \patchAmsMathEnvironmentForLineno{#1*}%
}
\AtBeginDocument{%
\patchBothAmsMathEnvironmentsForLineno{equation}%
\patchBothAmsMathEnvironmentsForLineno{align}%
\patchBothAmsMathEnvironmentsForLineno{flalign}%
\patchBothAmsMathEnvironmentsForLineno{alignat}%
\patchBothAmsMathEnvironmentsForLineno{gather}%
\patchBothAmsMathEnvironmentsForLineno{multline}%
\patchBothAmsMathEnvironmentsForLineno{eqnarray}%
}

\usepackage{hyperxmp}

\usepackage[pdftex,
            pdfauthor={\paperauthors},
            pdftitle={\paperasciititle},
            pdfkeywords={\paperkeywords},
            pdfcopyright={Copyright (C) \papercopyright},
            pdflicenseurl={\paperlicenceurl}]{hyperref}

\usepackage[all]{hypcap} 

\usepackage{xspace} 
\usepackage{upgreek}

\def\lhcb   {\mbox{LHCb}\xspace}

\def\MagUp {\mbox{\em Mag\kern -0.05em Up}\xspace}

\ifthenelse{\boolean{uprightparticles}}%
{

 \def\Ppi         {\ensuremath{\uppi}\xspace}

 \def\Ppsi        {\ensuremath{\uppsi}\xspace}

 \def\PDelta      {\ensuremath{\Delta}\xspace}                 
 \def\PXi         {\ensuremath{\Xi}\xspace}                 
 \def\PLambda     {\ensuremath{\Lambda}\xspace}                 
 \def\PSigma      {\ensuremath{\Sigma}\xspace}                 
 \def\POmega      {\ensuremath{\Omega}\xspace}                 
 \def\PUpsilon    {\ensuremath{\Upsilon}\xspace}

 \def\PB      {\ensuremath{\mathrm{B}}\xspace}                 
                  
 \def\PD      {\ensuremath{\mathrm{D}}\xspace}

 \def\PJ      {\ensuremath{\mathrm{J}}\xspace}                 
 \def\PK      {\ensuremath{\mathrm{K}}\xspace}

 \def\Pb      {\ensuremath{\mathrm{b}}\xspace}                 
 \def\Pc      {\ensuremath{\mathrm{c}}\xspace}

 \def\Pi      {\ensuremath{\mathrm{i}}\xspace}

 \def\Pp      {\ensuremath{\mathrm{p}}\xspace}

 \def\Ps      {\ensuremath{\mathrm{s}}\xspace}

 \def\thebaroffset{0.0em}
}
{

 \def\Ppi         {\ensuremath{\pi}\xspace}

 \def\Ppsi        {\ensuremath{\psi}\xspace}                 
                  
 \mathchardef\PDelta="7101
 \mathchardef\PXi="7104
 \mathchardef\PLambda="7103
 \mathchardef\PSigma="7106
 \mathchardef\POmega="710A
 \mathchardef\PUpsilon="7107
                  
 \def\PB      {\ensuremath{B}\xspace}                 
                  
 \def\PD      {\ensuremath{D}\xspace}

 \def\PJ      {\ensuremath{J}\xspace}                 
 \def\PK      {\ensuremath{K}\xspace}

 \def\Pb      {\ensuremath{b}\xspace}                 
 \def\Pc      {\ensuremath{c}\xspace}

 \def\Pi      {\ensuremath{i}\xspace}

 \def\Pp      {\ensuremath{p}\xspace}

 \def\Ps      {\ensuremath{s}\xspace}

 \def\thebaroffset{0.18em}
}
\newcommand{\offsetoverline}[2][\thebaroffset]{\kern #1\overline{\kern -#1 #2}}%

\makeatletter
\ifcase \@ptsize \relax
  \newcommand{\miniscule}{\@setfontsize\miniscule{4}{5}}
\or
  \newcommand{\miniscule}{\@setfontsize\miniscule{5}{6}}
\or
  \newcommand{\miniscule}{\@setfontsize\miniscule{5}{6}}
\fi
\makeatother

\DeclareRobustCommand{\optbar}[1]{\shortstack{{\miniscule (\rule[.5ex]{1.25em}{.18mm})}
  \\ [-.7ex] $#1$}}

\def\squark    {{\ensuremath{\Ps}}\xspace}

\def\cquark    {{\ensuremath{\Pc}}\xspace}

\def\bquark    {{\ensuremath{\Pb}}\xspace}

\def\pion   {{\ensuremath{\Ppi}}\xspace}
\def\piz    {{\ensuremath{\pion^0}}\xspace}
\def\pip    {{\ensuremath{\pion^+}}\xspace}
\def\pim    {{\ensuremath{\pion^-}}\xspace}
\def\pipm   {{\ensuremath{\pion^\pm}}\xspace}

\def\kaon    {{\ensuremath{\PK}}\xspace}
\def\Kbar    {{\ensuremath{\offsetoverline{\PK}}}\xspace}
\def\Kb      {{\ensuremath{\Kbar}}\xspace}
\def\KorKbar {\kern \thebaroffset\optbar{\kern -\thebaroffset \PK}{}\xspace}

\def\Kp      {{\ensuremath{\kaon^+}}\xspace}
\def\Km      {{\ensuremath{\kaon^-}}\xspace}

\def\D       {{\ensuremath{\PD}}\xspace}

\def\DorDbar {\kern \thebaroffset\optbar{\kern -\thebaroffset \PD}\xspace}
\def\Dz      {{\ensuremath{\D^0}}\xspace}

\def\Dp      {{\ensuremath{\D^+}}\xspace}

\def\Dstarp  {{\ensuremath{\D^{*+}}}\xspace}

\def\Dsp     {{\ensuremath{\D^+_\squark}}\xspace}

\def\B       {{\ensuremath{\PB}}\xspace}

\def\BorBbar {\kern \thebaroffset\optbar{\kern -\thebaroffset \PB}\xspace}

\def\Bd      {{\ensuremath{\B^0}}\xspace}

\def\BdorBdbar {\kern \thebaroffset\optbar{\kern -\thebaroffset \Bd}\xspace}
\def\Bu      {{\ensuremath{\B^+}}\xspace}

\def\Bp      {{\ensuremath{\Bu}}\xspace}

\def\Bs      {{\ensuremath{\B^0_\squark}}\xspace}

\def\BsorBsbar {\kern \thebaroffset\optbar{\kern -\thebaroffset \Bs}\xspace}

\def\jpsi     {{\ensuremath{{\PJ\mskip -3mu/\mskip -2mu\Ppsi\mskip 2mu}}}\xspace}

\def\Y#1S{\ensuremath{\PUpsilon{(#1S)}}\xspace}

\def\proton      {{\ensuremath{\Pp}}\xspace}

\def\Lz          {{\ensuremath{\PLambda}}\xspace}

\def\LorLbar     {\kern \thebaroffset\optbar{\kern -\thebaroffset \PLambda}\xspace}

\def\Sigmares    {{\ensuremath{\PSigma}}\xspace}

\def\Xires       {{\ensuremath{\PXi}}\xspace}

\def\Omegares    {{\ensuremath{\POmega}}\xspace}

\def\OmegabStarStar {{\ensuremath{\Omegares^{**-}_\bquark}}\xspace}

\def\Xic         {{\ensuremath{\Xires_\cquark}}\xspace}

\def\Xicp        {{\ensuremath{\Xires^+_\cquark}}\xspace}

\def\Omegac      {{\ensuremath{\Omegares^0_\cquark}}\xspace}

\def\Lb           {{\ensuremath{\Lz^0_\bquark}}\xspace}

\def\Xib          {{\ensuremath{\Xires_\bquark}}\xspace}
\def\Xibz         {{\ensuremath{\Xires^0_\bquark}}\xspace}
\def\Xibzprime    {{\ensuremath{\Xires^{\prime 0}_\bquark}}\xspace}
\def\Xibzstar    {{\ensuremath{\Xires^{*0}_\bquark}}\xspace}
\def\Xibmprime    {{\ensuremath{\Xires^{\prime -}_\bquark}}\xspace}
\def\Xibmstar    {{\ensuremath{\Xires^{*-}_\bquark}}\xspace}
\def\Xibm         {{\ensuremath{\Xires^-_\bquark}}\xspace}

\def\Omegab       {{\ensuremath{\Omegares^-_\bquark}}\xspace}

\def\OmegabA       {{\ensuremath{\Omegares_\bquark(6316)^-}}\xspace}
\def\OmegabB       {{\ensuremath{\Omegares_\bquark(6330)^-}}\xspace}
\def\OmegabC       {{\ensuremath{\Omegares_\bquark(6340)^-}}\xspace}
\def\OmegabD       {{\ensuremath{\Omegares_\bquark(6350)^-}}\xspace}

\newcommand{\decay}[2]{\ensuremath{#1\!\to #2}\xspace} 

\def\to                 {\ensuremath{\rightarrow}\xspace}

\def\AT#1     {\ensuremath{A_{\mathrm{T}}^{#1}}\xspace}           

\def\C#1      {\ensuremath{\mathcal{C}_{#1}}\xspace}                       
\def\Cp#1     {\ensuremath{\mathcal{C}_{#1}^{'}}\xspace}                    
\def\Ceff#1   {\ensuremath{\mathcal{C}_{#1}^{\mathrm{(eff)}}}\xspace}        
\def\Cpeff#1  {\ensuremath{\mathcal{C}_{#1}^{'\mathrm{(eff)}}}\xspace}       
\def\Ope#1    {\ensuremath{\mathcal{O}_{#1}}\xspace}                       
\def\Opep#1   {\ensuremath{\mathcal{O}_{#1}^{'}}\xspace}                    

\newcommand{\aunit}[1]{\ensuremath{\text{\,#1}}}       

\newcommand{\tev}{\aunit{Te\kern -0.1em V}\xspace}
\newcommand{\gev}{\aunit{Ge\kern -0.1em V}\xspace}
\newcommand{\mev}{\aunit{Me\kern -0.1em V}\xspace}
\newcommand{\kev}{\aunit{ke\kern -0.1em V}\xspace}
\newcommand{\ev}{\aunit{e\kern -0.1em V}\xspace}
\newcommand{\mevc}{\ensuremath{\aunit{Me\kern -0.1em V\!/}c}\xspace}
\newcommand{\gevc}{\ensuremath{\aunit{Ge\kern -0.1em V\!/}c}\xspace}
\newcommand{\mevcc}{\ensuremath{\aunit{Me\kern -0.1em V\!/}c^2}\xspace}
\newcommand{\gevcc}{\ensuremath{\aunit{Ge\kern -0.1em V\!/}c^2}\xspace}

\def\fb   {\ensuremath{\aunit{fb}}\xspace}
\def\invfb   {\ensuremath{\fb^{-1}}\xspace}

\newcommand{\chisq}{\ensuremath{\chi^2}\xspace}

\newcommand{\chisqip}{\ensuremath{\chi^2_{\text{IP}}}\xspace}

\def\gsim{{~\raise.15em\hbox{$>$}\kern-.85em
          \lower.35em\hbox{$\sim$}~}\xspace}
\def\lsim{{~\raise.15em\hbox{$<$}\kern-.85em
          \lower.35em\hbox{$\sim$}~}\xspace}

\def\tell1  {TELL1\xspace}
\def\ukl1   {UKL1\xspace}

\usepackage{cite} 
\usepackage{mciteplus}

\usepackage{longtable} 

\begin{document}

\renewcommand{\thefootnote}{\fnsymbol{footnote}}
\setcounter{footnote}{1}


\begin{titlepage}
\pagenumbering{roman}

\vspace*{-1.5cm}
\centerline{\large EUROPEAN ORGANIZATION FOR NUCLEAR RESEARCH (CERN)}
\vspace*{0.5cm}
\noindent
\begin{tabular*}{\linewidth}{lc@{\extracolsep{\fill}}r@{\extracolsep{0pt}}}
\ifthenelse{\boolean{pdflatex}}
{\vspace*{-1.5cm}\mbox{\!\!\!\includegraphics[width=.14\textwidth]{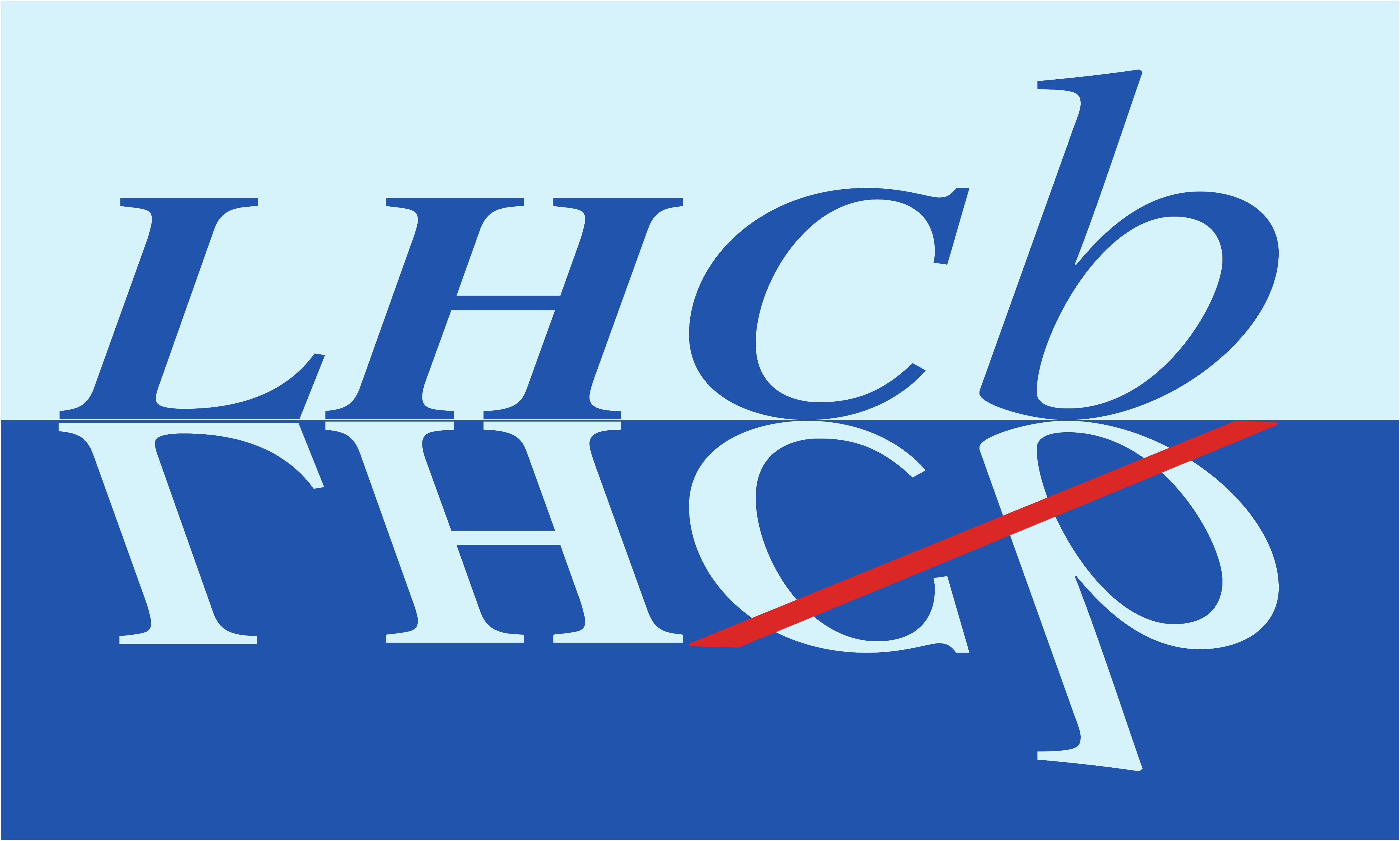}} & &}%
{\vspace*{-1.2cm}\mbox{\!\!\!\includegraphics[width=.12\textwidth]{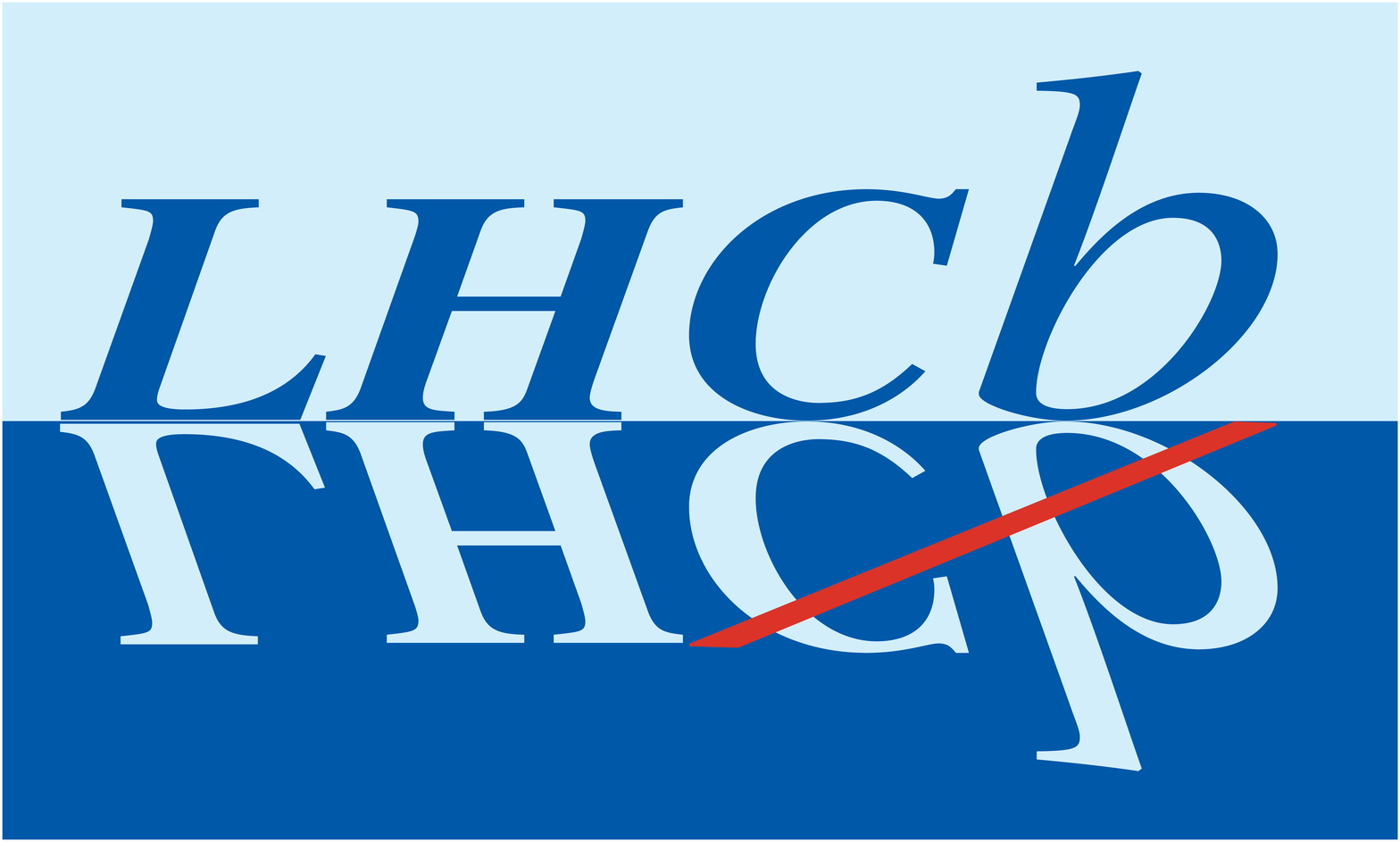}} & &}%
\\
 & & CERN-EP-2019-278 \\  
 & & LHCb-PAPER-2019-042 \\  
 & & Feb. 25, 2020 \\ 
 & & \\
\end{tabular*}

\vspace*{0.5cm}

{\normalfont\bfseries\boldmath\huge
\begin{center}
  \papertitle 
\end{center}
}

\vspace*{0.5cm}

\begin{center}
\paperauthors\footnote{Authors are listed at the end of this Letter.}
\end{center}

\vspace{\fill}

\begin{abstract}
  \noindent
We report four narrow peaks in the $\Xibz\Km$ mass spectrum obtained using $pp$ collisions at center-of-mass energies of 7, 8 and 13\tev, corresponding to a total integrated luminosity of 9\invfb recorded by the LHCb experiment. Referring to these states by their mass, the mass values are
\begin{align*}
m(\OmegabA) &= 6315.64\pm0.31\pm0.07\pm0.50\mev, \\
m(\OmegabB) &= 6330.30\pm0.28\pm0.07\pm0.50\mev, \\
m(\OmegabC) &= 6339.71\pm0.26\pm0.05\pm0.50\mev, \\
m(\OmegabD) &= 6349.88\pm0.35\pm0.05\pm0.50\mev,
\end{align*}
\noindent where the uncertainties are statistical, systematic and the last is due to the knowledge of the $\Xibz$ mass. The natural widths of
the three lower mass states are consistent with zero, and the 90\% confidence-level upper limits are determined to be ${\Gamma(\OmegabA)<2.8\mev}$, ${\Gamma(\OmegabB)<3.1\mev}$ and ${\Gamma(\OmegabC)<1.5\mev}$. The natural width of the $\OmegabD$ peak is
$1.4^{+1.0}_{-0.8}\pm0.1\mev$, which is 2.5$\sigma$ from zero and corresponds to an upper limit of $2.8\mev$. 
The peaks have local significances ranging from 3.6$\sigma$ to 7.2$\sigma$. 
After accounting for the look-elsewhere effect, the significances of the \OmegabA and \OmegabB
peaks are reduced to 2.1$\sigma$ and 2.6$\sigma$ respectively, while the two higher mass peaks exceed~5$\sigma$.
The observed peaks are consistent with expectations for excited $\Omegab$ resonances.
\end{abstract}

\vspace*{2.0cm}

\begin{center}
  Published in
  Phys.~Rev.~Lett. 
\end{center}

\vspace{\fill}

{\footnotesize 
\centerline{\copyright~\papercopyright. \href{\paperlicenceurl}{\paperlicence}.}}
\vspace*{2mm}

\end{titlepage}


\newpage
\setcounter{page}{2}
\mbox{~}
\cleardoublepage



\renewcommand{\thefootnote}{\arabic{footnote}}
\setcounter{footnote}{0}



\pagestyle{plain} 
\setcounter{page}{1}
\pagenumbering{arabic}

The study of hadrons containing heavy ($b$ or $c$) quarks has undergone a renaissance over the last couple of decades. During this time a plethora of new states have been observed, including candidates for four-quark (tetraquark) states, and more recently five-quark (pentaquark) states~\cite{LHCb-PAPER-2019-014,LHCb-PAPER-2015-029,LHCb-PAPER-2016-019}
(see Refs.~\cite{Brambilla:2019esw,Liu:2019zoy,Olsen:2017bmm} for recent reviews). In addition, a number of observations of peaking structures in the
invariant-mass spectra of final states containing $\Xicp\Km$~\cite{LHCb-PAPER-2017-002}, $\Xibz\pim$~\cite{LHCb-PAPER-2018-013}, 
$\Lb\pim$~\cite{LHCb-PAPER-2018-032}, and $\Lb\pip\pim$~\cite{LHCb-PAPER-2012-012,LHCb-PAPER-2019-025} have provided valuable experimental information to improve our understanding of quantum chromodynamics (QCD), the theory of the strong interaction.

Fueled by these observations, there has been a renewed interest in gaining a deeper 
theoretical understanding of hadronic structure. The constituent quark model~\cite{GellMann:1964nj,Zweig:352337,*Zweig:570209} has been very successful in describing the types of hadrons that form in nature and
how they fit into multiplets~\cite{PDG2018} based on the quantum numbers of the states.
While conventional baryons are understood to be states that contain three valence quarks, a deep understanding of how best
to describe these and other multi-quark states in terms of their fundamental constituents is still an open question. 
For example, in QCD, two quarks can exhibit attraction when in a $J^P=0^+$ quantum state, giving rise to
the notion that conventional baryons can be described as the bound state of a quark and a $qq^{\prime}$ diquark~\cite{Ida:1966ev,Lichtenberg:1967zz}.
These ideas are naturally extensible to describe tetraquark and pentaquark candidates~\cite{Brambilla:2019esw,Liu:2019zoy,Olsen:2017bmm}.

Recently, the LHCb experiment observed five narrow states, assumed to be excited $\Omegac$ baryons , which decay into $\Xicp\Km$~\cite{LHCb-PAPER-2017-002}.
These states have been analyzed from the perspective of constituent quark 
models and lattice QCD~\cite{Yoshida:2015tia,Garcilazo:2007eh,Thakkar:2016dna,Agaev:2017jyt,Agaev:2017ywp,Mao:2017wbz,Santopinto:2018ljf,Yang:2017qan,Roberts:2007ni,PhysRevD.98.076015,Cui_PhysRevD.99.094021,Wang_PhysRevD.99.014011,Shah:2018daf,PhysRevLett.119.042001,Zhao_PhysRevD.95.114024,Aliev:2018uby,Wang:2017kfr,PhysRevLett.119.042001}, quark-diquark 
models~\cite{Ebert:2007nw,Ebert:2011kk,Faustov:2018vgl,Qin:2019,Chen_PhysRevD.94.114016,Wang:2017vnc,Chen_PhysRevD.96.094015,Wang_PhysRevD.95.116010,Karliner:2017kfm,Wang2017,Wang2017R},
as well as molecular models~\cite{Chen:2017xat,Debastiani:2017ewu,Montana:2017kjw,Huang_PhysRevD.97.094013,Huang_PhysRevD.97.034027,Tolos:2018stp} and
pentaquark states~\cite{Kim:2018cku,Wang:2018alb,An:2017lwg}. Several of the
models that seek to describe these peaks also make predictions for $\Xibz\Km$ resonances. 
Since the quark contents of the $\Omegac$ and $\Omegab$ baryons are $css$ and $bss$, respectively, 
it is of great interest to search for analogous states in the $\Xibz\Km$ mass spectrum.

This Letter reports on a search for narrow resonances in the $\Xibz\Km$ mass spectrum close to the kinematic threshold. The search
uses data collected in $pp$ collisions with the LHCb detector at center-of-mass energies of 7, 8 and 13\tev, corresponding 
to integrated luminosities of 1, 2 and 6\invfb, respectively. Charge-conjugate processes are implicitly included,
and natural units with $\hbar=c=1$ are used throughout.

The \lhcb detector~\cite{LHCb-DP-2008-001,LHCb-DP-2014-002} is a single-arm forward
spectrometer covering the \mbox{pseudorapidity} range $2<\eta <5$,
designed for the study of particles containing \bquark or \cquark
quarks. Events are selected online by a trigger,
which consists of a hardware stage, based on information from the calorimeter and muon
systems, followed by a software stage, which applies a full event reconstruction~\cite{LHCb-DP-2012-004,BBDT}.
Simulated data samples are produced using the software packages described in 
Refs.~\cite{Sjostrand:2007gs,Sjostrand:2006za,LHCb-PROC-2010-056,Lange:2001uf,Golonka:2005pn,Allison:2006ve, *Agostinelli:2002hh,LHCb-PROC-2011-006}, and
are used to optimize selection requirements and to quantify the invariant-mass resolution of the LHCb detector.

Samples of $\Xibz$ candidates are formed by pairing $\Xicp$ and $\pim$ candidates, where
the $\Xicp$ decays are reconstructed in the $p\Km\pip$ final state. All final-state hadrons
must have particle-identification (PID) information consistent with the assigned particle hypothesis.
The final-state particles are also required to be inconsistent with originating from a primary $pp$ collision vertex (PV) by requiring that they have 
large $\chisqip$ with respect to all PVs in the event. The quantity $\chisqip$ is the difference 
in $\chisq$ of the vertex fit of a given PV when the particle (here $p$, $\Km$ or $\pip$) 
is included and excluded from the fit. 

\begin{figure}[tb]
\centering
\includegraphics[width=0.49\textwidth]{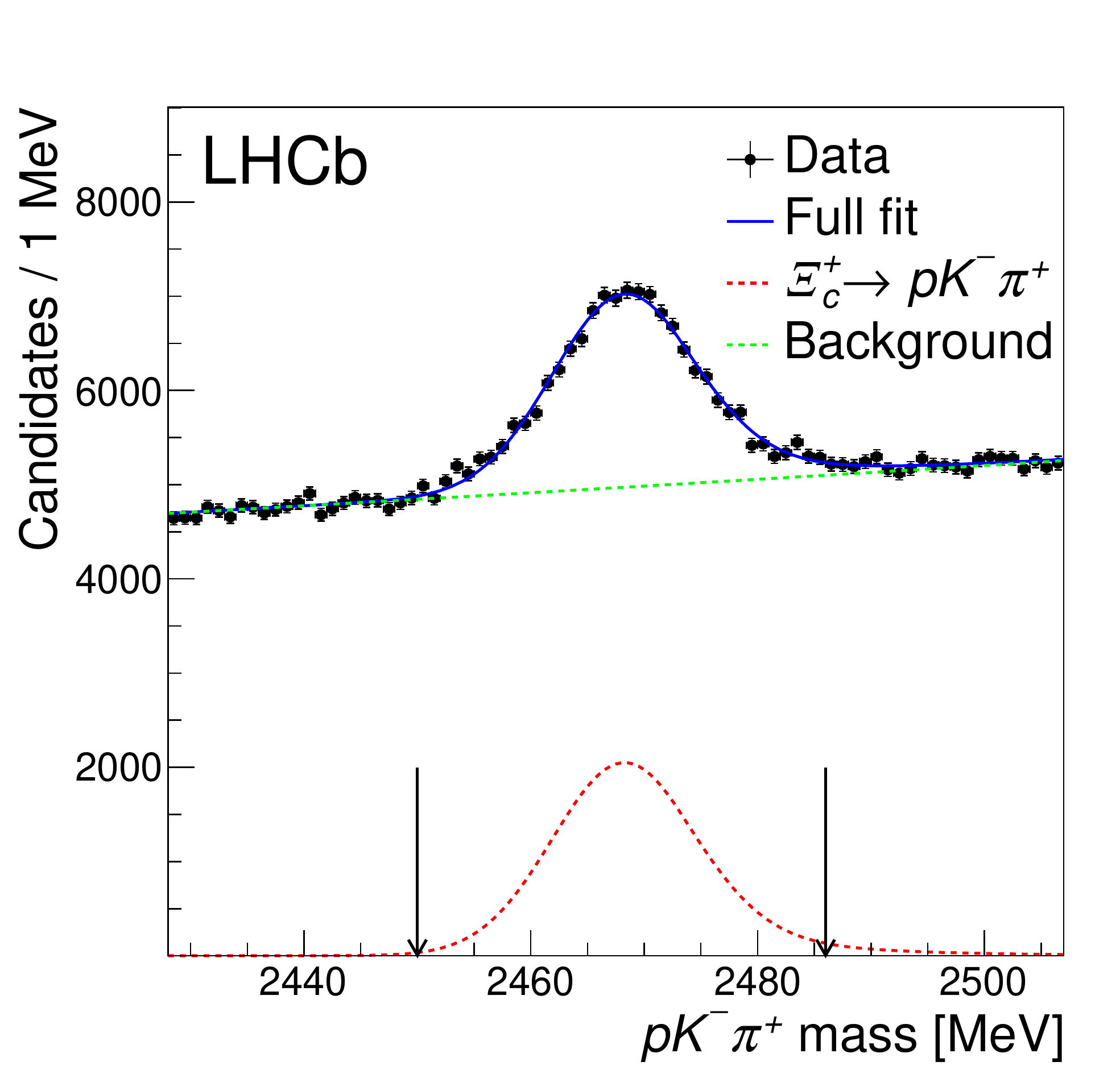}
\includegraphics[width=0.49\textwidth]{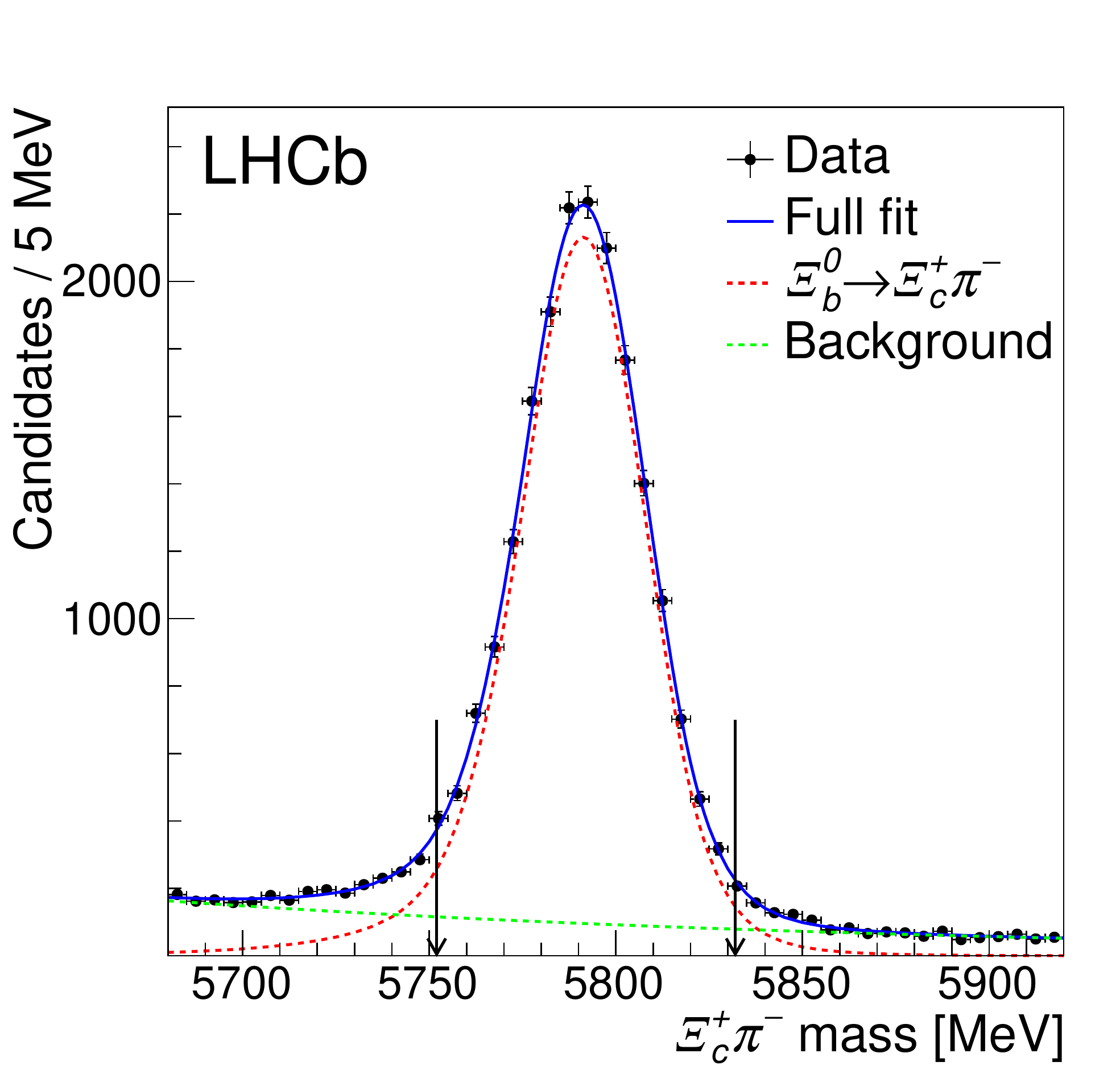}
\caption{\small{Invariant-mass spectrum for (left) $\Xicp\to p\Km\pip$ and (right) $\Xibz\to\Xicp\pim$ candidates in data 
passing the selection requirements described in the text. The arrows indicate the requirements on the invariant masses
that are applied in the subsequent stages of the analysis.}}
\label{fig:XibMass_All}
\end{figure}

The $\Xicp$ candidates must have a fitted vertex that is significantly
displaced from all PVs in the event and have an invariant mass within 18\mev of the
known $\Xicp$ mass~\cite{PDG2018}.  About 20\% of the $\Xicp$ background comprises misidentified
$\Dp\to\Km\pip\pip,~\Dp\to\Kp\Km\pip$, $\Dsp\to\Kp\Km\pip$ and $\Dstarp\to(\Dz\to\Km\pip)\pip$ decays, as well as misidentified $\phi$ mesons with $\phi\to\Kp\Km$ combined with an additional particle from elsewhere in the event.
These background contributions are removed by employing tighter PID requirements on candidates that are consistent
with any of these decay hypotheses, resulting in about 1\% loss of signal efficiency. The $p\Km\pip$ invariant-mass distribution
of $\Xicp$ candidates satisfying these selection requirements is shown in Fig.~\ref{fig:XibMass_All}\,(left).

The $\Xibz$ candidates are formed from $\Xicp\pim$ combinations that have a
significantly displaced decay vertex from all PVs in the event and a trajectory that is consistent with originating from one of them. The PV for which the $\Xibz$ candidate has the smallest $\chisqip$ is assigned to be the associated PV, and it is used subsequently to compute quantities such as the $\Xibz$ decay time. Candidates satisfying the requirement $5.6<M(\Xicp\pim)<6.0$\gev are retained, where $M$ designates the invariant mass of the system.

To further suppress background in the $\Xibz\to\Xicp\pim$ sample, a boosted decision tree (BDT) 
discriminant~\cite{AdaBoost} is used. The BDT exploits 21 input variables:
the decay times of the $\Xicp$ and $\Xibz$ candidates and the $\chisq$ values associated with their decay-vertex fits;
the angle between the $\Xibz$ momentum vector and the line that joins the $\Xibz$ decay vertex and its
associated PV; and for each final state particle the momentum, transverse momentum, $\chisqip$ and a PID response variable.
The PID response for final-state hadrons in the signal decay is obtained from large 
$\Dstarp\to(\Dz\to\Km\pip)\pip$ and $\Lz\to p\pim$ calibration samples in data~\cite{LHCb-DP-2012-003,LHCb-DP-2018-001}.
Simulated signal decays and background from the $\Xicp$ mass sidebands ($30<|M(p\Km\pip)-m_{\Xicp}|<50$\mev) in 
data are used to train the BDT, where $m$ refers to the mass of the indicated particle~\cite{PDG2018}. 
The chosen requirement on the BDT response provides a relative
signal efficiency of 90\%, and reduces the combinatorial background by about a factor of 2.5.
Overall, the offline selection requirements are about 75\% efficient on simulated decays, while reducing the background by about a factor of 40. 

Figure~\ref{fig:XibMass_All}\,(right) shows the $\Xicp\pim$ mass spectrum for candidates passing the 
above selection criteria. The spectrum is fit with the sum of two Crystal Ball~\cite{Skwarnicki:1986xj} functions with 
a common mean and opposite-side power-law tails to model the 
signal, and an exponential function to describe the background distribution. The fitted $\Xibz$ signal yield  
is $19\,200\pm200$. 

To search for peaking structures in the $\Xibz\Km$ mass spectrum, a requirement that 
$|M(\Xicp\pim)-m_{\Xibz}|<40\mev$ is imposed, which reduces the number of $\Xibz$ signal decays to about 18\,000.
Each $\Xibz$ candidate is combined with a $\Km$ candidate that is consistent with originating from a PV in the event.
The $\Xibz$ and $\Km$ trajectories are fitted to a common vertex, and that vertex is kinematically constrained to coincide with the PV associated with the $\Xibz$ candidate~\cite{Hulsbergen:2005pu}. The additional PV constraint improves the resolution
on the mass difference $\delta M\equiv M(\Xibz\Km)-M(\Xibz)$ by about a factor of two.

Random combinations of $\Xibz$ baryons with a $\Km$ candidate are the largest source of background in the $\Xibz\Km$ mass spectrum. 
To improve the expected signal-to-background ratio, a figure of merit, $\epsilon/(\sqrt{B}+5/2)$~\cite{Punzi:2003bu}, 
is used to optimize the requirements on the PID information of the $\Km$ candidates.
Here, $\epsilon$ is the efficiency as determined from simulation, and 
$B$ is the number of wrong-sign $\Xibz\Kp$ combinations in the region $520<\delta M<570$\mev passing the
PID requirement, scaled to a 10\mev mass window. The 10\mev width is chosen based on the search for
narrow peaks, since the low signal yields expected would make wide peaks difficult to separate from the
combinatorial background. 
The optimal requirement on the $\Km$ PID provides an efficiency of about 85\% and suppresses the background by a factor of about~2.5.

The decay of a resonance to $\Xibz\Km$ will produce peaks in the $\delta M$ spectrum. The experimental $\delta M$ resolution is obtained from simulated samples generated at several masses, $m_{\rm res}$. The resolution function is described by the sum of two Gaussian functions with a common mean. In addition, the width of the narrower Gaussian component, $\sigma_{\rm core}$, is fixed to be 45\% of that of the wider component, and its contribution is required to constitute 80\% of the total shape.
A smooth, monotonically increasing function, denoted as 
$\sigma(m_{\rm res})$, is then used to parameterize 
$\sigma_{\rm core}$ as a function of $m_{\rm res}$.
 In the $\delta M$ interval of interest, $\sigma(m_{\rm res})$ is in the range of 0.7--0.8\mev.

\begin{figure}[tb]
\centering
\includegraphics[width=0.98\textwidth]{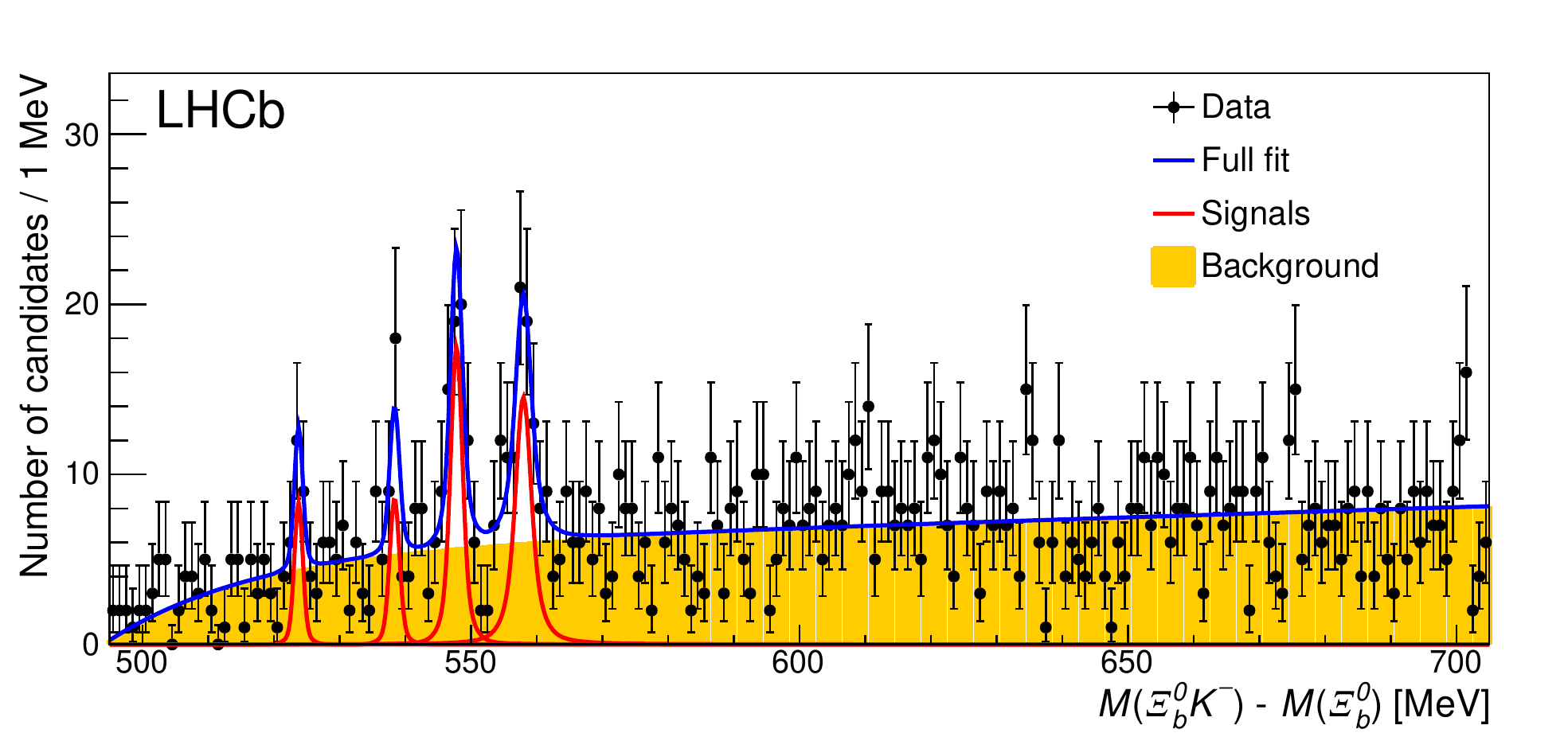}
\includegraphics[width=0.98\textwidth]{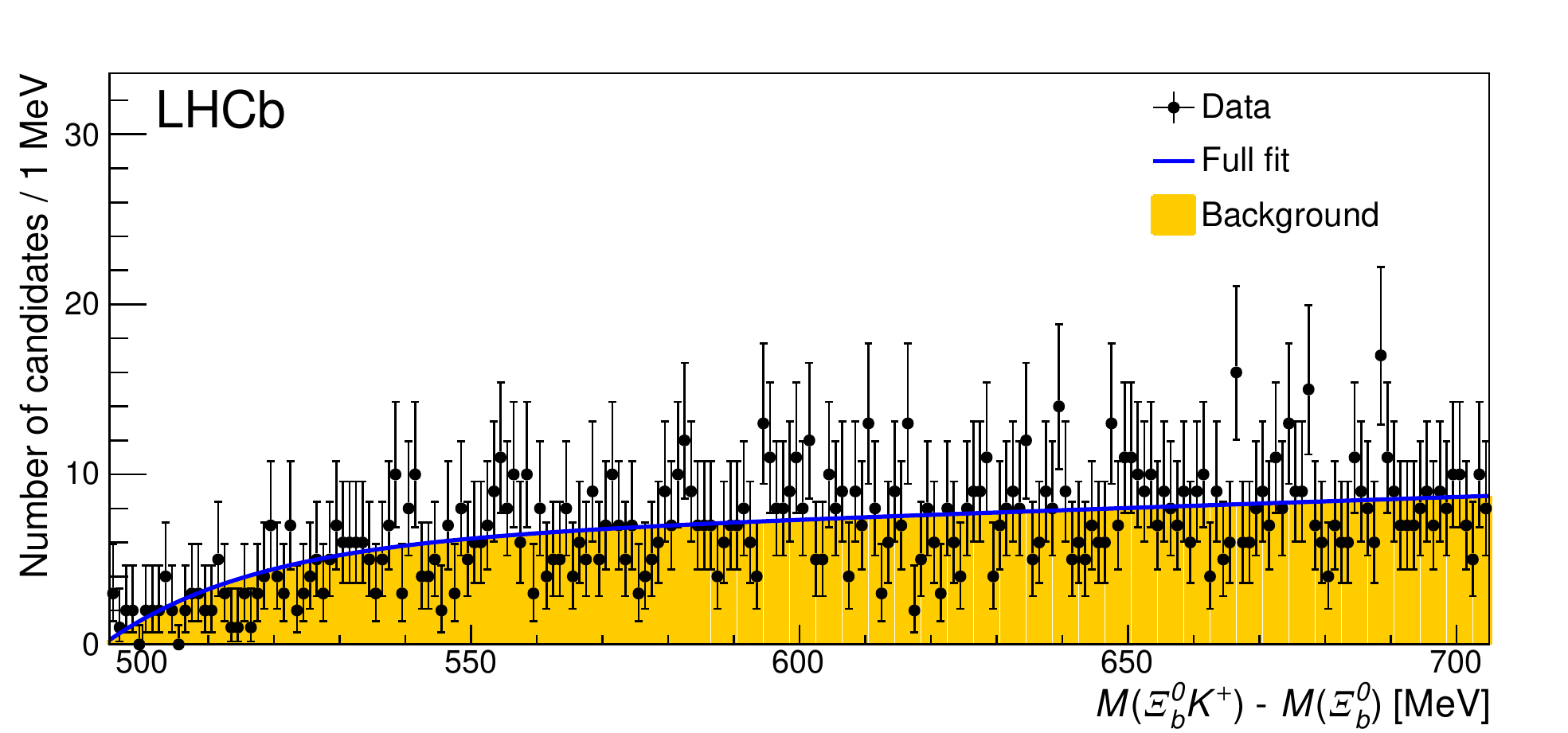}
\caption{\small{Distribution of the mass difference for (top)  right-sign $\Xibz\Km$ candidates, and (bottom) wrong-sign $\Xibz\Kp$ candidates, as described in the text.}}
\label{fig:OmegabStar}
\end{figure}

The $\delta M$ distributions for right-sign (RS) and wrong-sign (WS) candidates are shown in Fig.~\ref{fig:OmegabStar}, along with
fits to the spectra as described below. 
Four peaks are seen in the RS spectrum of $\Xibz\Km$ candidates (red curves), whereas no significant peaks are seen in the corresponding WS $\Xibz\Kp$ distribution.
To obtain the parameters of the peaks, a simultaneous unbinned extended 
maximum-likelihood fit is performed to the
RS and WS spectra. Each signal peak is described 
by an {\mbox{$S$--wave}} relativistic Breit--Wigner function~\cite{Jackson:1964zd} with a Blatt--Weisskopf 
barrier factor~\cite{Blatt}, convoluted with the resolution function $\sigma(m_{\rm res})$ described above. A common
background shape is used to describe both the RS and WS spectra, and is described by a smooth three-parameter monotonic
function that accounts for the $\Xibz\Km$ threshold. 

The peak values of $\delta M$, natural widths, signal yields, and the local and global significances are summarized in 
Table~\ref{tab:peakPars}. The local significance is obtained as ${{\mathcal{S}}_{\rm data}=\sqrt{2\log({{\mathcal{L}}_{\rm max}/\mathcal{L}}_0})}$, where 
${\mathcal{L}}_{\rm max}$ is the maximum value of the fit likelihood and ${\mathcal{L}}_0$ is the value obtained 
when a given peak's yield is fixed to zero. All peaks have natural width consistent with zero. The highest-mass peak has the largest width, which differs from zero by 2.5 standard deviations, as determined 
from a likelihood scan of the width parameter. 

To account for the look-elsewhere effect~\cite{Gross:2010qma}, which considers that the peak search extends over about a 200\mev wide mass region,
a large number of pseudoexperiments (pe) are generated. The pseudoexperiments use the nominal parameters from the fit to 
the data, with the signal yield of each peak, in turn, set to zero. The full mass region is
scanned in 0.5\mev steps to identify the most significant positive fluctuation outside of the region of 
the three retained peaks, from which the significance ${\mathcal{S}}_{\rm pe}$ is computed. From the corresponding distribution of ${\mathcal{S}}_{\rm pe}$ and the value ${\mathcal{S}}_{\rm data}$, 
a $p$-value --- expressed in Gaussian standard deviations --- is obtained for each peak, as shown in Table~\ref{tab:peakPars}. 

\begin{table*}[tb]
\begin{center}
\caption{\small{Peak positions, widths, signal yields, and local and global significances of the four mass peaks observed in the 
$\Xibz\Km$ mass spectra, as described in the text. The uncertainties are statistical only.}}
\begin{tabular}{ccccc}
\hline\hline
Peak of $\delta M$    & Width    &  Signal  &    \multicolumn{2}{c}{Significances~[$\sigma$]}  \\
   $[\!\mev]$             & $[\!\mev]$   &   yield  & Local  & Global  \\
\hline
\\ [-2.5ex]
$523.74\pm0.31$ & $0.00\,^{+\,0.7}_{-\,0.0}$  & $15\,_{-\,5}^{\,+6\phantom{0}}$  & 3.6 & 2.1 \\   
\\ [-2.5ex]
$538.40\pm0.28$ & $0.00\,^{+\,0.4}_{-\,0.0}$  & $18\,_{-\,5}^{+\,6\phantom{0}}$  & 3.7 & 2.6 \\  
\\ [-2.5ex]
$547.81\pm0.26$ & $0.47\,^{+\,0.6}_{-\,0.5}$  & $47\,_{-\,10}^{+\,11}$ & 7.2 & 6.7 \\   
\\ [-2.5ex]
$557.98\pm0.35$ & $1.4\,^{+\,1.0}_{-\,0.8}$  & $57\,_{-\,13}^{+\,14}$ & 7.0 & 6.2 \\   
\hline\hline
\end{tabular}
\label{tab:peakPars}
\end{center}
\end{table*}

The sources of systematic uncertainty that affect the measured masses are summarized in Table~\ref{tab:syst}. The momentum scale uncertainty 
is assessed by shifting the momentum scale of all charged tracks by $\pm$0.03\%~\cite{LHCb-PAPER-2013-011} in simulated decays, 
and evaluating the change in $\delta M$. The imperfect modeling of the energy loss in the detector material results in a
systematic uncertainty of 0.04\mev~\cite{LHCb-PAPER-2011-035}. 
The uncertainty due to the choice of signal model is assigned by fitting the data with an alternative
signal model composed of two Gaussian functions with a common mean. The largest change, 0.02\mev, is assigned as a systematic uncertainty to all of the peak positions. The background shape uncertainty is assessed by removing the influence of the WS data on the background shape, and fitting only the RS data; the difference in the peak positions with respect to the 
nominal fit is assigned as a systematic uncertainty. The relativistic Breit--Wigner signal shape in the nominal fit assumes that the decay proceeds through an $S$-wave, with an
interaction radius in the Blatt--Weisskopf barrier factor of $R=3\gev^{-1}$. Changing the angular momentum in the decay to $L=2$ ($D$-wave), 
and separately varying $R$ between 1 and 5~$\gev^{-1}$, leads to a negligible change in the peak positions. For the absolute mass determination, the world-average $\Xibz$ mass of $5791.9\pm0.5$\mev~\cite{PDG2018} is used.
The uncertainty of 0.5\mev on this mass dominates the systematic uncertainty and is quoted separately in the final results.
\begin{table*}[b]
\begin{center}
\caption{\small{Systematic uncertainties on the measured peak positions in the ${\delta M = M(\Xibz\Km)-M(\Xibz)}$ spectrum. The peaks are numbered in order of increasing mass.
}}
\begin{tabular}{ccccc}
\hline\hline
Source & Peak 1 & Peak 2 & Peak 3 & Peak 4 \\
       & [MeV] & [MeV] & [MeV] & [MeV] \\
\hline
Momentum scale        & 0.01 & 0.02 & 0.02 & 0.03 \\
Energy loss           & 0.04 & 0.04 & 0.04 & 0.04 \\
Signal shape          & 0.02  & 0.02  & 0.02 & 0.02 \\
Background            & 0.05 & 0.05 & 0.01 & 0.01 \\
\hline
Total                 & 0.07 & 0.07 & 0.05 & 0.05 \\  
\hline\hline
\end{tabular}
\label{tab:syst}
\end{center}
\end{table*}

The primary source of systematic uncertainty on the natural widths of the observed peaks is from an imperfect knowledge
of the $\delta M$ resolution, which is obtained from simulation. Based on previous studies of $\Dstarp\to\Dz\pip$ 
decays~\cite{LHCb-PAPER-2014-021}, the $\delta M$ resolution in simulation agrees with that of data within 10\%.
The impact of a $\pm$10\% variation in the resolution is evaluated using pseudoexperiments, where each experiment is
generated using the nominal signal resolution function, and fitted with a 10\% smaller or larger $\delta M$ resolution.
Deviations of $\pm$0.10\mev relative to the true value of the width are found for a range of input widths corresponding to that which is observed in data.
The upper limits on the natural width of the observed peaks are evaluated by convoluting the likelihoods with this 0.10\mev uncertainty, and finding the values of the widths that contain 90\% and 95\% of the integrated probability. For both the mass differences and widths, the total uncertainty is dominated by the statistical component.

\begin{table*}[tb]
\begin{center}
\caption{\small{Summary of the peak parameters of the four peaks, showing the peak positions of ${\delta M = M(\Xibz\Km)-M(\Xibz)}$, 
the masses, and 90\% (95\%) confidence level upper limits on the natural widths. The indicated uncertainties are statistical,
systematic, and due to the world-average value of the $\Xibz$ mass (for the masses). 
For the $\OmegabD$ peak, the central value of the width is also indicated.}}
\begin{tabular}{lccl}
\hline\hline
                      & $\delta M_{\rm peak}$ [MeV]    & Mass [MeV]   & Width [MeV] \\
\hline\\[-4mm]
$\OmegabA$ & $523.74\pm0.31\pm0.07$  & $6315.64\pm0.31\pm0.07\pm0.50$  & $<2.8~(4.2)$ \\[1.5mm]
$\OmegabB$ & $538.40\pm0.28\pm0.07$  & $6330.30\pm0.28\pm0.07\pm0.50$  & $<3.1~(4.7)$  \\[1.5mm]
$\OmegabC$ & $547.81\pm0.26\pm0.05$  & $6339.71\pm0.26\pm0.05\pm0.50$  &  $<1.5~(1.8)$ \\[1.5mm]
$\OmegabD$ & $557.98\pm0.35\pm0.05$  & $6349.88\pm0.35\pm0.05\pm0.50$  & $<2.8~(3.2)$ \\[0.2mm]
                &                                &                     &  $~~\,\,1.4\,^{+\,1.0}_{-\,0.8}\pm0.1$  \\[1mm]
\hline\hline
\end{tabular}
\label{tab:param_sum}
\end{center}
\end{table*}

The measured masses and widths of the four peaks in the $\Xibz\Km$  mass spectrum are summarized in Table~\ref{tab:param_sum}. They are qualitatively similar to those observed in the $\Xicp\Km$ mass spectrum~\cite{LHCb-PAPER-2017-002}.
Arguably, the simplest interpretation of these peaks is that they correspond to excited $\Omegab$ states, in particular the 
$L=1$ angular momentum excitations of the ground state, or possibly $n=2$ radial excitations. Many of the quark model calculations predict $L=1$ states in this mass 
region~\cite{Yoshida:2015tia,Garcilazo:2007eh,Thakkar:2016dna,Agaev:2017jyt,Agaev:2017ywp,Mao:2017wbz,Santopinto:2018ljf,Yang:2017qan,Roberts:2007ni,PhysRevD.98.076015,Wang_PhysRevD.99.014011,Wang:2017kfr}, and at least some of the states should be narrow~\cite{Agaev:2017ywp,Santopinto:2018ljf,Wang:2017kfr}. In particular, 
using the $^3P_0$ model, five states in this mass region are predicted, with approximately 8\mev mass splittings; 
the four lightest have partial width, $\Gamma(\Xibz\Km)$, below 1\mev, while that with the largest mass has
$\Gamma(\Xibz\Km)=1.49\mev$~\cite{Santopinto:2018ljf}. On the other hand, predictions
using the chiral quark-model indicate that the $J^P=\frac{3}{2}^-$ and $\frac{5}{2}^-$ states are narrow, but the $\frac{1}{2}^-$ states are wide, in the 50--100\mev range~\cite{Wang:2017kfr}.

Quark-diquark models have also predicted several excited $\Omegab$ states in the region around 6.3\gev~\cite{Ebert:2007nw,Ebert:2011kk,Karliner:2017kfm,PhysRevD.98.074032}, with mass splittings 
similar to those observed here. In an implementation of the $^3P_0$ model, the $J^P=\frac{3}{2}^-$ and $\frac{3}{2}^-$ are predicted to be narrow~\cite{PhysRevD.98.074032}.
Molecular models have also been employed, where two narrow $J^P=\frac{1}{2}^-$ states are predicted at 6405 and 6465\mev~\cite{Liang:2017ejq};
no statistically significant peaks are seen at those masses with the current dataset.

An alternate interpretation for one or more of the observed peaks is that they arise from the decay of a higher-mass excited $\Omegab$ state, $\OmegabStarStar\to\Xibzprime(\to\Xibz\piz)\Km$, where the $\piz$ meson is undetected. 
While the $\Xibmprime$, $\Xibmstar$~\cite{LHCb-PAPER-2014-021} and $\Xibzstar$~\cite{Chatrchyan:2012ni,LHCb-PAPER-2016-010} baryons have been observed,
the $\Xibzprime$ resonance is yet to be seen. If the $\Xibzprime$ mass is in the interval
$m_{\Xibz}+m_{\piz}<m_{\Xibzprime}<m_{\Xibmprime}$, each of the observed narrow peaks can be interpreted as
having originated from the above decay, provided that the corresponding $\OmegabStarStar$ state is narrow. In this case, their masses can be evaluated as
$m_{\OmegabStarStar}=m_{\Xibzprime}+\delta M_{\rm peak}$, where the values of $\delta M_{\rm peak}$ are taken from Table~\ref{tab:param_sum}. 
If the $\Xibzprime$ baryon can only decay electromagnetically to $\Xibz\gamma$, then the $\Xibz\Km$ peaks would be significantly broader and inconsistent with our data. 

In summary, $pp$ collision data collected with the LHCb experiment at center-of-mass energies of $7$, $8$ and $13\tev$, corresponding to
integrated luminosities of 1, 2 and 6\invfb, respectively, have been used to search for near-threshold $\Xibz\Km$ resonances. Four new peaks are seen. 
Two of the peaks, the $\OmegabC$ and $\OmegabD$, are observed with global (local) significance of 6.7 (7.2) and 6.2 (7.0), respectively, while the
two lower-mass peaks have global (local) significance of 2.1 (3.6) and 2.6 (3.7). The peaks are consistent with expectations for excited $\Omegab$ resonances.


\section*{Acknowledgements}
%
%
\noindent We express our gratitude to our colleagues in the CERN
accelerator departments for the excellent performance of the LHC. We
thank the technical and administrative staff at the LHCb
institutes.
We acknowledge support from CERN and from the national agencies:
CAPES, CNPq, FAPERJ and FINEP (Brazil); 
MOST and NSFC (China); 
CNRS/IN2P3 (France); 
BMBF, DFG and MPG (Germany); 
INFN (Italy); 
NWO (Netherlands); 
MNiSW and NCN (Poland); 
MEN/IFA (Romania); 
MSHE (Russia); 
MinECo (Spain); 
SNSF and SER (Switzerland); 
NASU (Ukraine); 
STFC (United Kingdom); 
DOE NP and NSF (USA).
We acknowledge the computing resources that are provided by CERN, IN2P3
(France), KIT and DESY (Germany), INFN (Italy), SURF (Netherlands),
PIC (Spain), GridPP (United Kingdom), RRCKI and Yandex
LLC (Russia), CSCS (Switzerland), IFIN-HH (Romania), CBPF (Brazil),
PL-GRID (Poland) and OSC (USA).
We are indebted to the communities behind the multiple open-source
software packages on which we depend.
Individual groups or members have received support from
AvH Foundation (Germany);
EPLANET, Marie Sk\l{}odowska-Curie Actions and ERC (European Union);
ANR, Labex P2IO and OCEVU, and R\'{e}gion Auvergne-Rh\^{o}ne-Alpes (France);
Key Research Program of Frontier Sciences of CAS, CAS PIFI, and the Thousand Talents Program (China);
RFBR, RSF and Yandex LLC (Russia);
GVA, XuntaGal and GENCAT (Spain);
the Royal Society
and the Leverhulme Trust (United Kingdom).


\clearpage
\newpage

\ifx\mcitethebibliography\mciteundefinedmacro
\PackageError{LHCb.bst}{mciteplus.sty has not been loaded}
{This bibstyle requires the use of the mciteplus package.}\fi
\providecommand{\href}[2]{#2}


\newpage

\centerline
{\large\bf LHCb collaboration}
\begin
{flushleft}
\small
R.~Aaij$^{31}$,
C.~Abell{\'a}n~Beteta$^{49}$,
T.~Ackernley$^{59}$,
B.~Adeva$^{45}$,
M.~Adinolfi$^{53}$,
H.~Afsharnia$^{9}$,
C.A.~Aidala$^{80}$,
S.~Aiola$^{25}$,
Z.~Ajaltouni$^{9}$,
S.~Akar$^{66}$
% ...

{\footnotesize \it

% ...
$ ^{9}$Universit{\'e} Clermont Auvergne, CNRS/IN2P3, LPC, Clermont-Ferrand, France\\
% ...
$ ^{25}$INFN Sezione di Milano, Milano, Italy\\
% ...
$ ^{31}$Nikhef National Institute for Subatomic Physics, Amsterdam, Netherlands\\
% ...
$ ^{45}$Instituto Galego de F{\'\i}sica de Altas Enerx{\'\i}as (IGFAE), Universidade de Santiago de Compostela, Santiago de Compostela, Spain\\
% ...
$ ^{49}$Physik-Institut, Universit{\"a}t Z{\"u}rich, Z{\"u}rich, Switzerland\\
% ...
$ ^{53}$H.H. Wills Physics Laboratory, University of Bristol, Bristol, United Kingdom\\
% ...
$ ^{59}$Oliver Lodge Laboratory, University of Liverpool, Liverpool, United Kingdom\\
% ...
$ ^{66}$Los Alamos National Laboratory (LANL), Los Alamos, United States\\
$ ^{67}$Syracuse University, Syracuse, NY, United States\\
% ...
$ ^{80}$University of Michigan, Ann Arbor, United States, associated to $^{67}$\\
\bigskip
% ...
}
\end{flushleft}

\end{document}